\documentclass{desyproc}

\begin{document}
\title{WISP Dark Matter eXperiment and Prospects for Broadband Dark Matter Searches in the 1\,$\mu$eV--10\,meV Mass Range}


\author{{\slshape 
Dieter Horns$^1$, Axel Lindner$^2$, Andrei Lobanov$^{3,1,\dag}$,
Andreas Ringwald$^2$}\\[1ex]
$^1$Institut f\"ur Experimentalphysik, Universit\"at Hamburg, Germany \\
$^2$Deutsches Elektronen-Synchrotron (DESY), Hamburg, Germany\\
$^3$Max-Planck-Institut f\"ur Radioastronomie, Bonn, Germany\\[0.5ex]
$^\dag$corresponding author.
}

\contribID{lobanov_andrei}

\desyproc{DESY-PROC-2014-XX}
\acronym{Patras 2014} 
\doi  

\maketitle

\begin{abstract}

  Light cold dark matter consisting of weakly interacting slim
  (or sub-eV) particles (WISPs) has been in the focus of a large number of
  studies made over the past two decades.  The QCD axion and
  axion-like particles with masses in the 0.1\,$\mu$eV--100\,meV are
  strong candidates for the dark matter particle, together with hidden
  photons with masses below $\lesssim 100$\,meV.  This motivates
  several new initiatives in the field, including the WISP Dark Matter
  eXperiment (WISPDMX) and novel conceptual approaches for broad-band
  WISP searches using radiometry measurements in large volume
  chambers. First results and future prospects for these experiments
  are discussed in this contribution.
\end{abstract}

\section{WISP dark matter searches in the 1\,$\mu$eV--10\,meV mass range}

Searches for light cold dark matter (DM) consisting of weakly
interacting slim particles
(WISPs)~\cite{Essig:2013es,Jaeckel:2010ni,Ringwald:2012hr,Hewett:2012if}
are gaining prominence, with a number of experiments conducted and
proposed detecting hidden photons (HPs), QCD axions, and axion-like
particles (ALPs) with
astrophysical~\cite{Mirizzi:2009iz,Mirizzi:2009nq,Chelouche:2009,Pshirkov:2009,Horns:2012mm,Horns:2012me,Lobanov:2013pr}
and laboratory
measurements~\cite{Bradley:2003rv,Asztalos:2009yp,Wagner:2010mi,Baum:2013ba,Horns:2013,Horns:2013pa}.

Best revealed by their coupling to standard model (SM) photons, 
WISPs can be non-thermally produced in the early Universe
\cite{Sikivie:2008ax} and may give rise to dark matter for a broad
range of the particle mass and the photon coupling
strength~\cite{Arias:2012az,Redondo:2013lm}.  The photon coupling
of axions/ALPs, $g_{\mathrm{a}\gamma} \propto 1/f_\mathrm{a}$, depends
on the energy scale $f_\mathrm{a}$ of the symmetry breaking
responsible for the given particle, while the hidden photon coupling,
$\chi \propto g_\mathrm{h}$, is determined by the hidden gauge
coupling, $g_\mathrm{h}$~\cite{Arias:2012az}.

At particle masses above $\sim$$10^{-3}$\,eV, the existing constraints
effectively rule out WISPs as DM particles, while cosmologically
viable lower ranges of particle reach down to
$\sim$$10^{-6}$--$10^{-7}$\,eV for axions/ALPs and may extend to much
lower masses with for HP and ALP, as well as with ``fine tuning'' of
the axion models.  The mass range $10^{-7}$--$10^{-3}$\,eV corresponds
to the radio regime at frequencies of 24\,MHz--240\,GHz where highly
sensitive measurement techniques are developed for radioastronomical
measurements, with typical detection levels of $\lesssim 10^{-22}$\,W.

\section{Narrowband experiments}

The most sensitive laboratory HP and axion/ALP DM searches performed
so far in the mass range $10^{-7}$--$10^{-3}$\,eV have utilised the
``haloscope'' approach~\cite{Sikivie:1983pr} which employs resonant
microwave cavities lowering substantially the detection threshold in a
narrow band around each of the cavity
resonances~\cite{Bradley:2003rv,Asztalos:2009yp,Wagner:2010mi,DePanfilis:1987dk,Wuensch:1989sa,Hagmann:1990tj}. The
WISP DM signal results from conversion of DM halo particles inside the
cavity volume, $V$, into normal photons.  This conversion is achieved
via spontaneous kinetic mixing of HP with SM photons or via the Primakov
process induced by an external magnetic field, ${\mathbf B}$, for axions/ALPs.

The output power of the axion/ALP conversion signal is $P_\mathrm{out}
\propto G\,Q\,V\,|{\mathbf B}|^2\,$, where $Q$ is the resonant
enhancement factor (quality factor) of microwave cavity and $G$ is
the fraction of cavity volume (form factor) in which the electric
field of the converted photon can be
detected~\cite{Asztalos:2001tf,Baker:2011ar}.  Both $G$ and $Q$ depend
on the cavity design and the resonant mode employed in the
measurement. The respective $P_\mathrm{out}$ for the HP signal does
not have a dependence on ${\mathbf B}$.

The fractional bandwidth of a haloscope measurement is $\propto
Q^{-1}$, with typical values of $Q$ not exceeding $\sim 10^5$.  The
WISP DM signal itself is restricted to within a fractional bandwidth
of $Q_\mathrm{DM}^{-1} \approx \sigma_\mathrm{DM}^2/c^2 \approx
10^{-6}$, with $\sigma_\mathrm{DM}$ describing the velocity dispersion
of the dark matter halo. A measurement made at a frequency, $\nu$, and
lasting for a time, $t$, can detect the WISP signal at an $\mathrm{SNR}
\propto \sqrt{t\,/\,W}\,P_\mathrm{out}/(k_\mathrm{b}\,T_\mathrm{n})$,
where $W$ is the signal bandwidth, $T_\mathrm{n}$ is the noise
temperature of the detector, and $k_\mathrm{b}$ is the Boltzmann
constant.

The exceptional sensitivity of microwave cavity experiments comes at
the expense of rather low scanning speeds of $\lesssim
1$\,GHz/year~\cite{Bradley:2003rv,Hagmann:1989hu}, which makes it
difficult to implement this kind of measurements for scanning over
large ranges of particle mass. 
To overcome this difficulty, new experimental concepts are being
developed that could relax the necessity of using the resonant
enhancement and working in a radiometer mode with an effective
$Q=1$~\cite{Horns:2013,Horns:2013pa,Jaeckel:2013pr,Jaeckel:2013jc,Doebrich:2014pt}.

\begin{figure}[t!]
\begin{center}
\includegraphics[width=0.8\textwidth]{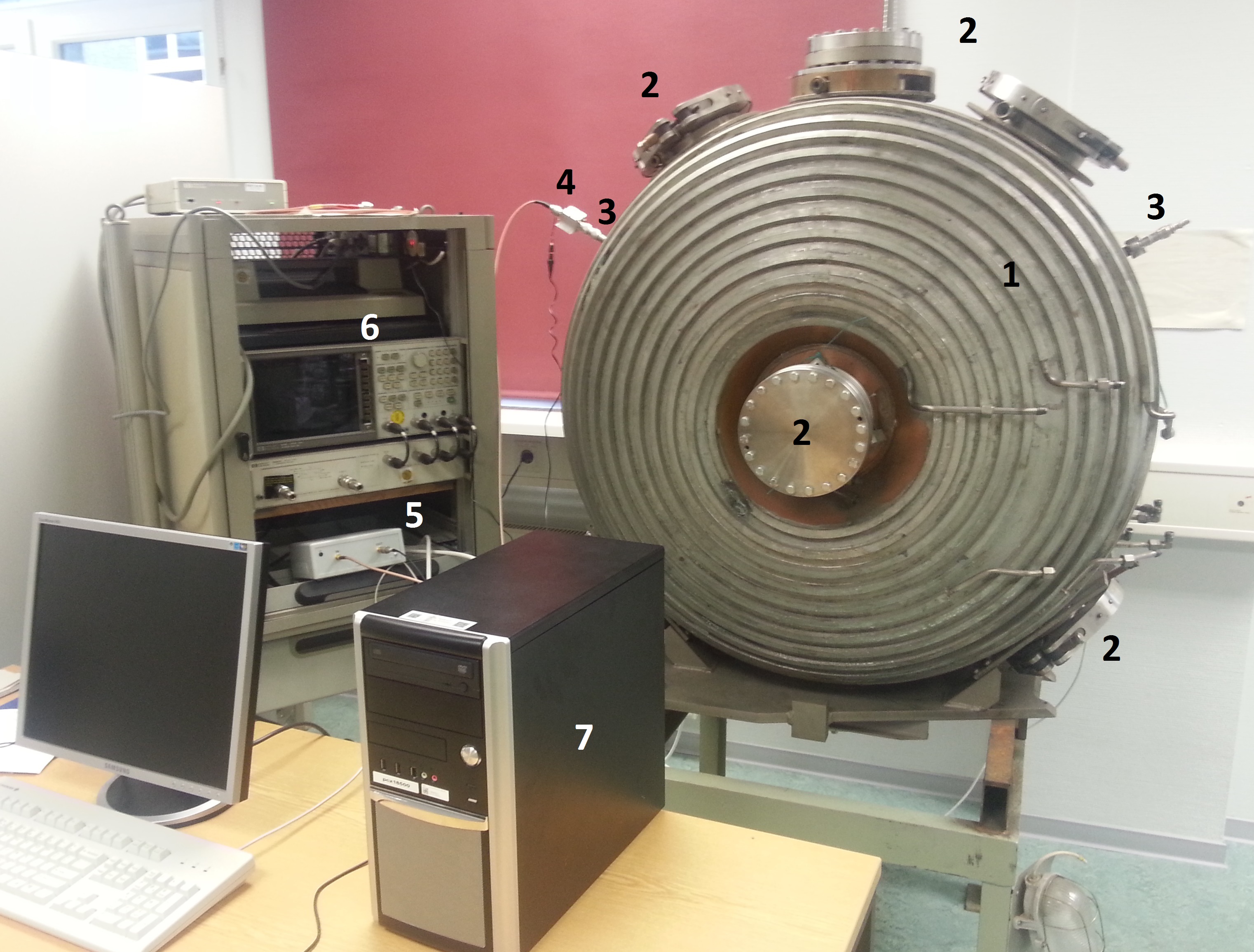}
\end{center}
\caption{WISPDMX setup for the initial measurements made at nominal
  resonant frequencies of the microwave cavity. Indicated with numbers
  are: 1 -- a 208-MHz microwave cavity of the type used in the proton
  ring of the DESY HERA collider; 2 - cavity ports (two on the axis
  of the cavity and five on the radial wall of the oblate cylinder; 3
  -- antenna ports for two magnetic-loop antennas inserted in the
  cavity; 4 -- first stage amplifier (WantCom, 0.1-1.0 GHz, 22.5 dB
  gain, 0.6 dB noise figure); 5 -- second stage amplifier (MITEQ,
  0.1-0.9 GHz, 18.2 gain, 1.0 dB noise figure); 6 -- network analyzer
  (HP 85047A); 7 -- control computer, with a 12-bit digitizer Alazar
  ATS-9360 operated via a PCIe Gen2 interface.}
\label{fg:lobanov1}
\end{figure}

\subsection{WISP Dark Matter eXperiment}

At frequencies below 1\,GHz, the ADMX
experiment\cite{Bradley:2003rv,Asztalos:2009yp,Wagner:2010mi} has
employed the haloscope approach to probe the HP and axion/ALP dark matter
in the 460--860\,MHz (1.9--3.6\,$\mu$eV) range~\cite{Asztalos:2010pr},
and a high frequency extension, ADMX-HF is planned for he 4--40\,GHz
frequency range~\cite{vanBibber:2013ca,Rosenberg:2014pa}. The WISP Dark Matter
eXperiment (WISPDMX, Fig.~\ref{fg:lobanov1}) extends the haloscope
searches to particle masses below 1.9\,$\mu$eV, aiming to cover the
range 200--600\,MHz (0.8--2.5\,$\mu$eV).

WISPDMX utilises a 208\,MHz resonant cavity of the type used at the
proton accelerator ring of the DESY HERA
collider~\cite{Gamp:1990pa}.  The cavity has a volume of 460 liters
and a nominal resonant amplification factor $Q=55000$ at the ground
TM$_{010}$ mode. The signal is amplified by a broad-band 0.2--1\,GHz
amplifier chain with a total gain of 40\,dB. Broad-band digitization
and FFT analysis of the signal are performed using a commercial 12-bit
spectral analyzer, enabling simultaneous measurements at several
resonant modes at frequencies of up to 600\,MHz.

The main specific aspects of WISPDMX measurements are: 1)~broadband
recording in the 200--600\,MHz band at a resolution of $\le
150$\,Hz, 2)~the use of multiple resonant modes tuned by a plunger
assembly consisting of two plungers, and 3)~the planned use of rotation
of the cavity in the magnetic field in order to enable axion
measurements at multiple resonant modes as well.

The WISPDMX measurements are split into three different stages: 1)~HP DM
searches at the nominal frequencies of the resonant modes of the HERA
cavity; 2)~HP DM searches with cavity tuning (covering up to 70\% of
the 200--600\,MHz band); and 3)~axion/ALP DM searches using the DESY
H1 solenoid magnet which provides $B=1.15$\,T in a volume of
7.2\,m$^3$.

\begin{figure}[ht!]
\begin{center}
\includegraphics[width=1.0\textwidth]{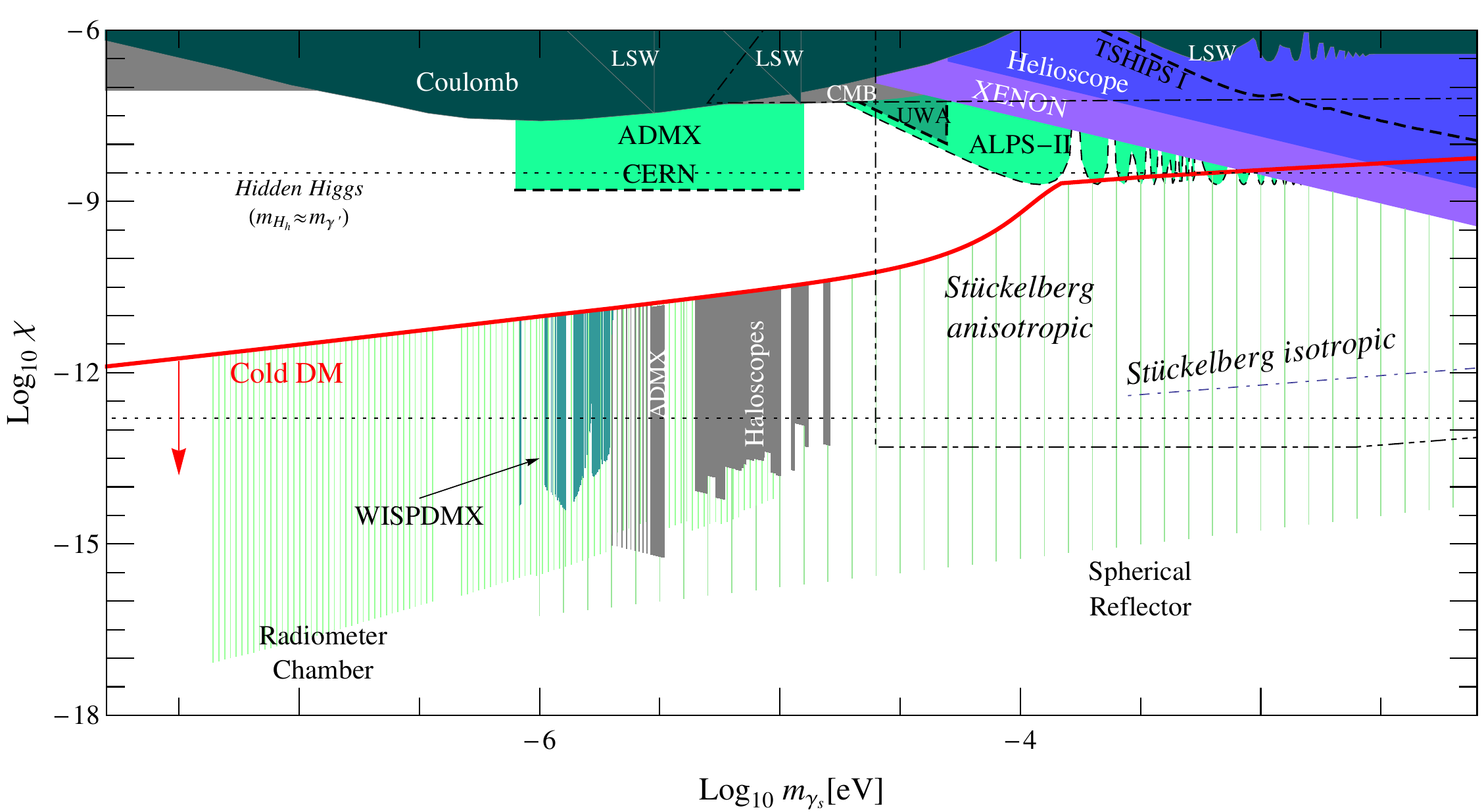}
\end{center}
\caption{Exclusion limits for hidden photon coupling to normal photons
  from existing (white captions) and planned (black captions)
  experiments (adapted from~\cite{Hewett:2012if}). Theoretical
  expectations for the hidden photon DM are indicated in red. The
  expected sensitivity limits of the second phase of the WISPDMX
  experiment are marked with turquoise color, reaching well into the DM
  favored range of photon coupling. The WISPDMX sensitivity is
  calculated for a setup with two tuning plungers operating
  simultaneously. The hatched areas illustrate the expected
  sensitivity of broadband experiments using the spherical reflector
  and the radiometry chamber approaches.}
\label{fg:lobanov2}
\end{figure}

\begin{figure}[t!]
\begin{center}
\includegraphics[width=1.0\textwidth]{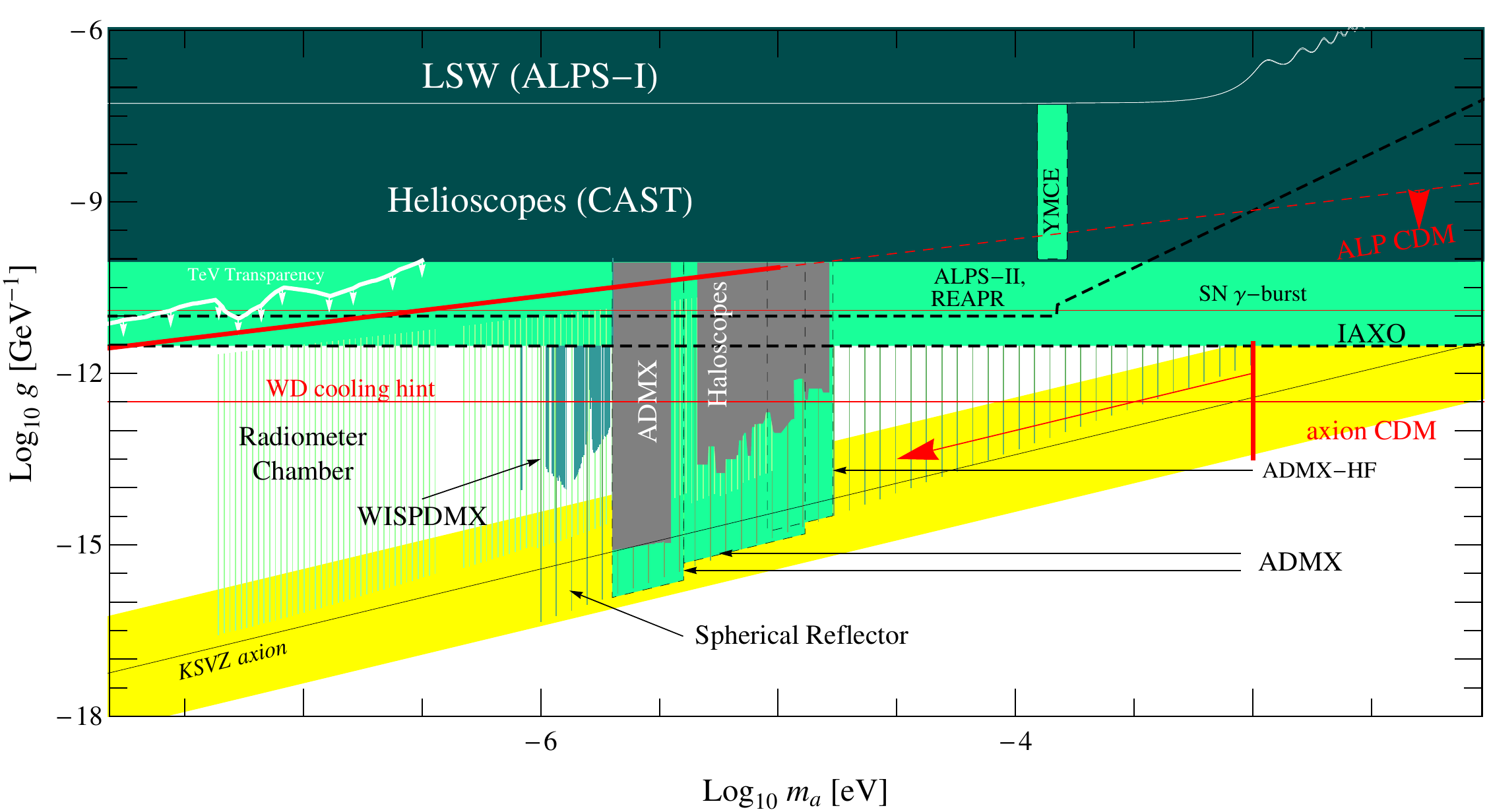}
\end{center}
\caption{ Exclusion limits for axion/ALP photon coupling to normal
  photons from existing (white captions) and planned (black captions)
  experiments (adapted from~\cite{Hewett:2012if}). The yellow band
  corresponds to the parameter space allowed for the QCD
  axion. Theoretical expectations for the axion/ALP DM are indicated
  in red. The expected sensitivity limits of the second phase of the
  WISPDMX experiment are marked with turquoise color, reaching well
  into the DM favored range of ALP coupling. The WISPDMX sensitivity
  is calculated for a setup with two tuning plungers operating
  simultaneously. The hatched areas illustrate the expected
  sensitivity of broadband experiments using the spherical reflector
  and the radiometry chamber approaches.}
\label{fg:lobanov3}
\end{figure}

The first phase of the WISPDMX measurements has been
completed~\cite{Baum:2013ba}, and the final results are being prepared
for publication. The WISPDMX setup used for these measurements is
shown in Fig.~\ref{fg:lobanov1}. Table~\ref{tb:lobanov1} lists
preliminary exclusion limits obtained for five different resonant
modes of the cavity. These limits reach below $\chi = 3\times
10^{-12}$ at all of the modes used in the measurement, and already
these limits probe the parameter space admitting HP DM. The
expected exclusion limits for the HP and axion/ALP DM searches in the
next two phases of WISPDMX are shown in
Figs.~\ref{fg:lobanov2}-\ref{fg:lobanov3}.

\begin{table}[t!]
\begin{center}
\caption{WISPDMX measurements at nominal resonant modes in the 200-600\,MHz range} 
\medskip
\begin{tabular}{l|cccccc}\hline\hline
Mode & $\nu$ & $Q$ & ${\cal G}$ & $P_\mathrm{det}$ & $m_{\gamma_\mathrm{s}}$ & \multicolumn{1}{c}{$\chi$} \\
 & [MHz] &  &  & [$\times$$10^{-14}$W] & [$\mu$eV] & \multicolumn{1}{c}{[$\times$$10^{-13}$]} \\ \hline
TM$_{010}$ & 207.87961& 55405& 0.429& $1.08$& 0.85972093 &$17.0$ \\
TE$_{111}$ & 321.45113& 59770& 0.674& $1.08$& 1.3294150  &~$8.4$ \\
TE$_{111}$ & 322.74845& 58900& 0.671& $1.08$& 1.3347803  &~$8.5$ \\
TM$_{020}$ & 454.42411& 44340& 0.317& $1.08$& 1.8793470  &$10.1$ \\
TE$_{112}$ & 510.62681& 71597& 0.020& $1.09$& 2.1117827  &$28.2$ \\
TE$_{112}$ & 515.97110& 67840& 0.019& $1.09$& 2.1338849  &$29.5$ \\
TE$_{120}$ & 577.59175& 60350& 0.036& $1.10$& 2.3887274  &$20.4$ \\
TE$_{120}$ & 579.25126& 66520& 0.037& $1.10$& 2.3955906  &$19.1$ \\\hline
\end{tabular}
\end{center}
{\small {\bf Column designation:}~resonant mode of the cavity, with its respective resonant frequency, $\nu$, quality factor, $Q$, geometrical form factor, ${\cal G}$, measured noise power, $P_\mathrm{det}$, hidden photon mass, $m_{\gamma_\mathrm{s}}$, and 95\% exclusion limit, $\chi$, for hidden photon coupling to normal photons, calculated for the antenna coupling $\kappa = 0.01$~\cite{Baum:2013ba}.}
\label{tb:lobanov1}
\end{table}

\section{Broadband experiments}

The relatively low scanning speed ({\em e.g.}, $\approx 100$\,MHz/yr
for WISPDMX and $\sim 1$\,GHz/yr for ADMX) and limited tunability of
resonant experiments make them difficult to be used for covering the
entire 1\,$\mu$eV--10\,meV range of particle mass. The gain in
sensitivity provided by the resonant enhancement, $Q$, needs to be
factored against the bandwidth reduction of the order of $Q^{-1}$,
which may result in a somewhat counterintuitive conclusion that
non-resonant, broadband approaches may be preferred when dealing with
a mass range spanning over several decades.

Indeed, for axion/ALP searches, reaching a desired sensitivity to the
photon coupling implies a measurement time $t\propto T^2_\mathrm{sys}
B^{-4} V^{-2} G^{-2} Q^{-2}$. A broadband measurement, with
$Q_\mathrm{b}\equiv 1$, needs to last $Q^2$ times longer to reach the
same sensitivity. However, the broadband measurement probes the entire
mass range at once, while it would require a (potentially very large)
number of measurements, $N_\mathrm{mes} = 1 +
\log(\alpha)\,/\,\log[Q/(Q-1)]$, in order to cover a range of
particle mass from $m_1$ to $m_2 = \alpha\,m_1$ (with $\alpha > 1$).
Then, the broadband
approach would be more efficient in case if $t_\mathrm{b} <
t_\mathrm{n}\,N_\mathrm{mes}$. This corresponds to the following
comparison between these two types of measurement:
\[
1 + Q \log\,\alpha > 
\left(\frac{T_\mathrm{b}}{T_\mathrm{n}}\right)^{2}
\left(\frac{B_\mathrm{b}}{B_\mathrm{n}}\right)^{-4}
\left(\frac{V_\mathrm{b}}{V_\mathrm{n}}\right)^{-2}
\left(\frac{G_\mathrm{b}}{G_\mathrm{n}}\right)^{-2}\,,
\]
where the subscripts $b$ and $n$ refer to the respective parameters of
the broadband and narrowband measurements. For typical experimental
settings, one can expect that $T_\mathrm{b} \sim 100\, T_\mathrm{n}$,
$B_\mathrm{b} \sim B_\mathrm{n}$, $V_\mathrm{b} \sim 100
V_\mathrm{n}$, and $G_\mathrm{b} \sim 0.01 G_\mathrm{n}$. In such
conditions, a narrowband experiment with $Q>10^4/\log\,\alpha$ would
be less efficient in scanning over the given range of mass.  Applied to
the entire 0.1\,$\mu$eV--10\,meV range of mass, this condition implies
restricting the effective resonance enhancement of narrowband
measurements to less than $\sim 2000$, which effectively disfavors the
narrowband approach for addressing such a large range of particle
mass.

The measurement bandwidth of radiometry experiments is
limited only by the detector technology, with modern detectors
employed in radio astronomy routinely providing bandwidths in excess
of 1 GHz and spectral resolutions of better than $10^{6}$.

One possibility for a radiometer experiment is to employ a spherical
dish reflector that provides a signal enhancement proportional to the
area of the
reflector~\cite{Horns:2013,Jaeckel:2013pr,Jaeckel:2013jc,Doebrich:2014pt}.
This option is well-suited for making measurements at higher
frequencies (shorter wavelengths, $\lambda$), as the effective signal
enhancement is $\propto A_\mathrm{dish}/\lambda^2$, where
$A_\mathrm{dish}$ is the reflecting area~\cite{Horns:2013}. It is
therefore expected that this concept would be best applicable at
$\lambda \lesssim 1$\,cm ($\nu \gtrsim 30$\,GHz).

At lower frequencies, another attractive possibility for engaging into
the broadband measurements is to use the combination of large chamber
volume and strong magnetic field provided by superconducting TOKAMAKs
or stellarators such as the Wendelstein 7-X
stellarator~\cite{Hirsch:2008cp} in Greifswald (providing $B=3\,T$ in
a 30\,m$^3$ volume).  

\subsection{Radiometry chamber experiments}

More generally, the stellarator approach
signifies a conceptual shift to employing a large, magnetized volume of
space which can be probed in a radiometry mode ({\em radiometer
  chamber}), without resorting to a resonant enhancement of the
signal.  The exclusion limits expected to be achievable with the
spherical reflector experiments and with the measurements made in a
radiometer chamber with the volume and magnetic field similar to those
of the Wendelstein stellarator are shown in
Fig.~\ref{fg:lobanov2}-\ref{fg:lobanov3}.

Deriving from the stellarator approach, a large chamber can be
designed specifically for the radiometer searches, with the inner
walls of the chamber covered by multiple fractal antenna elements
providing a broad-band receiving response in the 0.1-25\,GHz frequency
range with a high efficiency and a nearly homogeneous azimuthal
receiving pattern~\cite{Azari:2008fr,Khan:2008pf}. Combination of
multiple receiving elements provides also the possibility for achieve
directional sensitivity to the incoming photons. 

The directional sensitivity can be realized through high time
resolution enabling phase difference measurements between individual
fractal antenna elements. Making such measurements is well within the
reach of the present day data recording implemented for radio
interferometric experiments with the recording rate as high as 16
Gigabit/sec~\cite{Whitney:2013mk}. This recording rate provides a time
resolution down to 0.12 nanoseconds, thus enabling phase measurements
over effective travel path difference $\delta_\mathrm{l} \approx 4$
millimetres.  This corresponds to an angular resolution
$\phi_\mathrm{p}$ in the range of $ R/(L/\delta_\mathrm{l}+1)
[(R^2+L^2)^{1/2} - \delta_\mathrm{l}]^{-1} \lesssim \phi_\mathrm{p}
\lesssim (2 R/\delta_\mathrm{l}+1)^{1/2}/(R/\delta_\mathrm{l}+1)$,
where $R$ and $L$ denote the smallest and largest dimensions of the
chamber, respectively. For a chamber fitted into the HERA H1 magnet
(assuming $R \approx 1$\,m and $L \approx 3$\,m), this corresponds to
angular resolutions of $0.05^{\circ} \lesssim \phi_\mathrm{p} \lesssim
5^{\circ}$.  The radiometry chamber measurements could therefore
provide an attractive experimental concept for performing WISP DM
searches in the range of particle mass between 0.1 and 2\,$\mu$eV,
thus closing the low end of the parameter space still open for the
axion dark matter.

\section*{Acknowledgments}

AL acknowledges support from the Collaborative Research Center
(Sonderforschungsbereich) SFB 676 ``Particles, Strings, and the Early
Universe'' funded by the German Research Society (Deutsche
Forschungsgemeinschaft, DFG).  WISPDMX is performed in a
collaboration between the University of Hamburg, DESY, and
Max-Planck-Institut f\"ur Radioastronomie in Bonn. WISPDMX has been
supported through an SFB 676 lump sum grant and a PIER 'Ideen Fonds'
grant.


\begin{footnotesize}

\end{footnotesize}


\end{document}